\documentclass[11pt]{article}
\usepackage{amsmath}
\usepackage{amsfonts}
\usepackage{amssymb}
\usepackage{theorem}

\def\Th{\Theta}
\def\l{\lambda}
\def\p{\partial}
\newtheorem{theorem}{Proposition}

\newcommand{\be}{\begin{equation}}
\newcommand{\ee}{\end{equation}}
\newcommand{\bea}{\begin{eqnarray}}
\newcommand{\eea}{\end{eqnarray}}
\newcommand{\beaa}{\begin{eqnarray*}}
\newcommand{\eeaa}{\end{eqnarray*}}

\newcommand{\nn}{\nonumber}
\renewcommand{\d}{\mathrm{d}}
\begin{document}
\title{On the dressing method for Dunajski
anti-self-duality equation}
\author{
L.V. Bogdanov\thanks
{L.D. Landau ITP, Kosygin str. 2,
Moscow 119334, Russia, e-mail
leonid@landau.ac.ru},
V.S. Dryuma\thanks
{IMI AS RM, Academy str,5,
MD2001, Kishinev, Moldova},
S.V. Manakov\thanks
{L.D. Landau ITP, Kosygin str. 2,
Moscow 119334, Russia}
}
\maketitle
\begin{abstract}
A dressing scheme applicable to Dunajski
equation is developed. Simple example of constructing solutions
in terms of implicit functions is considered. Dunajski equation hierarchy
is described,
its Lax-Sato form is presented.
Dunajsky equation hierarchy is characterised by
conservation of three-dimensional volume form, in which a spectral
variable is taken into account. Some reductions of Dunajsky equation hierarchy,
including waterbag-type reduction,
are studied.
\end{abstract}
\section{Dunajski equation}
Dunajski equation \cite{Dun} is a representative of the class integrable sytems
arising in the context of complex relativity \cite{Plebanski}-\cite{dun3}. 
It is closely connected
to the selebrated Pleba\'nski heavenly equations \cite{Plebanski}
and in some sense generalizes them.
It describes anti-self-dual null-K\"ahler structures. In \cite{Dun}
it was shown that all null-K\"ahler metrics (signature (2,2)) locally admit a canonical
Pleba\'nski form
\be
\label{metric}
g=\d w\d x+\d z \d y-\Th_{xx}\d z^2-\Th_{yy}\d w^2+2\Th_{xy}\d w\d z.
\ee
The conformal anti-self-duality (ASD) condition leads to Dunajski equation
\be
\label{nk1}
\Th_{wx}+\Th_{zy}+\Th_{xx}\Th_{yy}-\Th_{xy}^2=f,
\ee
\be
\label{nk2}
\square f=f_{xw}+f_{yz}+
\Th_{yy}f_{xx}+\Th_{xx}f_{yy}-2\Th_{xy}f_{xy}=0.
\ee

Equations (\ref{nk1},\ref{nk2}) represent a compatibility condition
for the linear system $L_0\Psi=L_1\Psi=0$, where
$\Psi=\Psi(w, z, x, y, \l)$
and
\begin{eqnarray}
\label{Lax2}
L_0&=&(\p_w-\Th_{xy}\p_y+\Th_{yy} \p_x)-\l\p_y+f_y\p_{\l},\nonumber\\
L_1&=&(\p_z+\Th_{xx}\p_y-\Th_{xy} \p_x)+\l\p_x-f_x\p_{\l}.
\end{eqnarray}

The case $f=0$ corresponds to metrics of the form (\ref{metric})
satisfying Einstein equations,
and Dunajski equation (\ref{nk1}), (\ref{nk2}) reduces to Pleba\'nski second
heavenly equation \cite{Plebanski}.

 In the case
 \[
 \Theta=-1/6\,{ a_1}(w,z){x}^{3}+1/2\,{ a_2}(w,z){x}^{2}y-
1/2\,{ a_3}(w,z)x{y}^{2}+\]\[+1/6\,{ a_4}(w,z){y}^{3}
\]
the geodesics of the metrics (\ref{metric}) are determined by
solutions of projectively flat second order ODE
\[
{\frac {d^{2}}{d{z}^{2}}}w(z)+{ a_1}(w,z)\left ({\frac
{d}{dz}}w(z) \right )^{3}+3\,{ a_2}(w,z)\left ({\frac
{d}{dz}}w(z)\right )^{2}+\]\[+3 \,{ a_3}(w,z){\frac {d}{dz}}w(z)+{
a_4}(w,z)=0.
\]
As an example \cite{Dryum}, it is possible to consider equation with the
coefficients 
\[
a_1=2\exp(\phi(w,z)),~ a_2=-\phi_z,~ a_3=\phi_w,~
a_4=-2\exp(\phi(w,z)),
\]
where $\phi(w,z)$ is solution of the Wylczynski-Tzitzeika equation
\[
\phi_{w z}=4\exp(2\phi)-\exp(-\phi).
\]

In this paper we develop methods of
integrable systems theory for Dunajski equation, using the ideas
of the works \cite{MS1}-\cite{HEred} and also 
\cite{Takasaki,Takasaki1}. We present the results  
without many technical details, which will be described in a more extensive text.
\section{Dressing scheme}
Let us consider nonlinear vector  Riemann problem of the form
\bea
\mathbf{S}_+= \mathbf{F}(\mathbf{S}_-),
\label{Riemann}
\eea
where $\mathbf{S}_+$, $\mathbf{S}_-$ denote the boundary values
of the N-component vector function on the sides of some
oriented curve $\gamma$ in the complex plane of the variable
$\lambda$. The problem is to find the function analitic
outside the curve with some fixed singularities at infinity which satisfies
(\ref{Riemann}). This problem is connected with a class of integrable equations,
which can be represented as a commutation relation for vector fields containing
a derivative on the spectral variable.

The more specific setting relevant for Dunajski equation is the following.
We consider three-component problem (\ref{Riemann})
for the functions
\beaa
&&
S^0\rightarrow\lambda+O(\frac{1}{\lambda}),
\\&&
S^1\rightarrow -\lambda z+x+O(\frac{1}{\lambda}),
\\&&
S^2\rightarrow \lambda w+y+O(\frac{1}{\lambda}),\quad \l\rightarrow \infty,
\eeaa
where $x,y,w,z$ are the variables of Dunajski equation (`times').
We suggest that for given $\mathbf F$ solution of the problem
($\ref{Riemann}$) exists and is unique (at least locally
in $x,y,w,z$).

Let us consider a linearized problem
\beaa
\delta S^i_+=\sum_j F^i_{,j}\delta S^j_-.
\eeaa
The linear space of solutions of this problem is spanned by the functions
$\mathbf{S}_x$, $\mathbf{S}_y$, $\mathbf{S}_\lambda$,
which can be multiplied by arbitrary function of spectral variable and times.
The presence of $\mathbf{S}_\lambda$ in the basis is the main difference
between the dressing schemes for the heavenly equation \cite{MS1,heav} and
Dunajski equation.

Expanding the functions $\mathbf{S}_z$, $\mathbf{S}_w$ into the basis,
we obtain linear equations
\begin{eqnarray}
\label{Lax3}
((\p_w+u_{y}\p_y+v_{y}\p_x)-\l\p_y+f_{y}\p_{\l})\mathbf{S}=0,
\nonumber\\
((\p_z-u_{x}\p_y-v_{x}\p_x)+\l\p_x-f_{y}\p_{\l})\mathbf{S}=0,
\end{eqnarray}
where $u$, $v$, $f$ can be expressed through the coefficients of expansion
of $S^0$, $S^1$, $S^2$
at $\lambda=\infty$,
\bea
&&
u=S^2_1-wS^0_1,\quad v=S^1_1+zS^0_1, \quad f=S^0_1,
\label{uvf}
\\&&
S^0=\lambda+
\sum_{n=1}^\infty\frac{S^0_n}{\lambda^n},\;
S^1=-z\l+x+
\sum_{n=1}^\infty\frac{S^1_n}{\lambda^n},\;
\nn\\&&
S^2=w\l+y+
\sum_{n=1}^\infty\frac{S^2_n}{\lambda^n},\quad \l=\infty.
\nn
\eea
To get a Lax pair for Dunajski equation, we shold consider the reduction
$
v_y=-u_x
$,
then we can inroduce a potential $\Th$,
\be
v=\Th_x,\; u=-\Th_y.
\label{uvtheta}
\ee
\begin{theorem}
Sufficient condition to provide the reduction
$$
v_y=-u_x
$$
in terms of the Riemann problem
(\ref{Riemann}) is
\be
\det F^i_{,j}=1.
\label{Rvol}
\ee
\end{theorem}
\subsection{An example}
Now will consider a simple example of constructing solution to Dunajski equation
using the problem (\ref{Riemann}). We use the problem of the form
\bea
&&
S^1_+=S^1_-,
\label{ex1}
\\&&
S^2_+=S^2_-F^{-1}(S^2_-\cdot S^0_-,S^1_-),
\label{ex2}
\\&&
S^0_+=S^0_-F(S^2_-\cdot S^0_-,S^1_-),
\label{ex3}
\eea
where $F$ is an arbitrary function of two variables.
It is easy to check that the reduction condition (\ref{Rvol}) is indeed
satisfied in this case.
Equation (\ref{ex1}) implies that $S^1=-\lambda z+x$. Substituting
this solution to linear
equations (\ref{Lax3}) (or using (\ref{uvf})), we obtain $v=zf$.

The second important property of the problem we use is that the function
$S^2\cdot S^0$ is analytic. Then we get an expression
\be
\phi=S^2\cdot S^0=\l^2 w+\l y+2fw +u.
\label{phi}
\ee
Equation (\ref{ex3}) now reads
\beaa
S^0_+=S^0_-F(\phi,-\lambda z+x).
\eeaa
The solution to this equation looks like
\beaa
S^0=\lambda\exp\left(\frac{1}{2\pi\text{i}}
\int_\gamma\frac{\d \l'}{\l-\l'}\ln F(\phi(\l'),-\lambda' z+x)
\right)
\eeaa
Considering the expansion of this expression in $\lambda$, we obtain
the equations
\bea
&&
\frac{1}{2\pi\text{i}}
\int_\gamma{\d \l}\ln F(\phi(\l),-\lambda z+x)=0,
\\&&
\frac{1}{2\pi\text{i}}
\int_\gamma\frac{\d \l}{\l}\ln F(\phi(\l),-\lambda z+x)=f.
\eea
Taking into account expression (\ref{phi}), we come to the conclusion
that these equations define the functions $u$, $f$ as implicit functions.
Solution to Dunajski equation is then defined by the relation
$$u=-\Th_y.$$
Thus we have obtained a solution to Dunajski equation, depending on
arbitrary function of two variables, in terms of implicit functions.

Functional dependence on the function of two variables indicates
that the solution we have constructed correspons to some
(2+1)-dimensional reduction of Dunajski equation. It is possible
to find the reduced equations explicitely, using the fact that linear
equations (\ref{Lax2}) have  analytic solutions $\phi$ and $-\lambda z+x$.
Substituting these solutions to (\ref{Lax2}) and using (\ref{uvtheta}),
we obtain
\bea
&&
(\p_w-\Th_{xy}\p_y+\Th_{yy} \p_x)(2wf-\Th_y)+yf_y=0,
\label{firstred}
\\&&
(\p_z+\Th_{xx}\p_y-\Th_{xy} \p_x)(2wf-\Th_y)-yf_x=0,\nn
\\&&
zf=\Theta_x.\nn
\eea
\section{Dunajski equation hierarchy}
The framework developed here is closely coonected with the framework of
of hyper-K\"ahler hierarchy developed by
Takasaki \cite{Takasaki,Takasaki1},
see also \cite{heav, HEred}. Though there are some essential differences
(the volume form is used instead of symplectic form, the spectral variable is included
to the form), the technique and ideas of the proofs are very similar.
Here we omit the details, planning to present them later.

To define Dunajski equation hierarchy, we consider three formal
Laurent series in $\lambda$, depending on two infinite sets of
additional variables (`times')
\beaa
&&
S^0=\lambda+\sum_{n=1}^\infty S^0_n(\mathbf{t}^1,\mathbf{t}^2)\l^{-n},
\\&&
S^1=\sum_{n=0}^\infty t^1_n (S^0)^{n}+
\sum_{n=1}^\infty S^1_n(\mathbf{t}^1,\mathbf{t}^2)\l^{-n}
\\&&
S^2=\sum_{n=0}^\infty t^2_n (S^0)^{n}+
\sum_{n=1}^\infty S^2_n(\mathbf{t}^1,\mathbf{t}^2)\l^{-n},
\eeaa
We denote $x=t^1_0$, $y=t^2_0$,
$
\mathbf{S}=
\left(
\begin{array}{c}
S^1\\
S^2
\end{array}
\right),
$
$\partial^1_n=\frac{\partial}{\partial t^1_n}$,
$\partial^2_n=\frac{\partial}{\partial t^2_n}$ and
introduce  the projectors $(\sum_{-\infty}^{\infty}u_n z^n)_+
=\sum_{n=0}^{\infty}u_n z^n$,
$(\sum_{-\infty}^{\infty}u_n z^n)_-=\sum_{-\infty}^{n=-1}u_n z^n$.

Dunajski equation hierarchy is defined by the relation
\be
(\d S^0\wedge \d S^1\wedge \d S^2)_-=0,
\label{analyticity0}
\ee
where the differential includes both times and a
spectral variable,
\beaa
\d f=\sum_{n=0}^{\infty}\partial^1_n f \d t^1_n +
\sum_{n=0}^{\infty}\partial^2_n f \d t^2_n
+ \partial_\l f \d \l.
\eeaa
This is a crucial difference with the heavenly equation hierarchy,
where only the times are taken into account.
Relation (\ref{analyticity0})
plays a role similar to the role of the famous Hirota bilinear identity for KP hierarchy.
This relation is equivalent to the Lax-Sato form of
Dunajski equation hierarchy.
\begin{theorem}
The relation
(\ref{analyticity0})
is equivalent to the set of equations
\bea
&&
\partial^1_n\mathbf{S}=\sum_{i=0,1,2}(J^{-1}_{1i} (S^0)^n)_+
{\partial_i}\mathbf{S},
\label{Dun1}
\\
&&
\partial^2_n\mathbf{S}= \sum_{i=0,1,2}(J^{-1}_{2i} (S^0)^n)_+
{\partial_i}\mathbf{S},
\label{Dun2}
\\
&&
\det J=1,
\label{volume}
\eea
where
\be
J=
\begin{pmatrix}
S^0_\l&S^1_\l&S^2_\l\\
S^0_x&S^1_x&S^2_x\\
S^0_y&S^1_y&S^2_y
\end{pmatrix},
\label{J}
\ee
$\partial_0=\partial_\lambda$, $\partial_1=\partial_x$, $\partial_2=\partial_y$.
\label{formDun}
\end{theorem}

\textbf{Remark} Formula (\ref{volume}) defines  a reduction for equations
(\ref{Dun1}, \ref{Dun2}).
The general hierarchy in the unreduced case is given by equations
(\ref{Dun1}, \ref{Dun2}),
and the analogue of relation (\ref{analyticity0}) is
\be
((\det J)^{-1}\d S^0\wedge \d S^1\wedge \d S^2)_-=0.
\label{analyticity00}
\ee

In a more explicit form, Dunajski equation hierarchy
(\ref{Dun1}, \ref{Dun2}) can be written as
\bea
\partial^1_n\mathbf{S}=
\left(
(S^0)^{n}
\begin{vmatrix}
S^0_\l & S^2_\l\\
S^0_y & S^2_y
\end{vmatrix}
\right)_+\partial_x \mathbf{S}-
\left(
(S^0)^{n}
\begin{vmatrix}
S^0_\l & S^2_\l\\
S^0_x & S^2_x
\end{vmatrix}
\right)_+\partial_y \mathbf{S}+
\nn\\
\left(
(S^0)^{n}
\begin{vmatrix}
S^0_x & S^2_x\\
S^0_y & S^2_y
\end{vmatrix}
\right)_+\partial_\l \mathbf{S}
,
\label{Dun11}
\eea
\bea
\partial^2_n\mathbf{S}=
-\left(
(S^0)^{n}
\begin{vmatrix}
S^0_\l & S^1_\l\\
S^0_y & S^1_y
\end{vmatrix}
\right)_+\partial_x \mathbf{S}+
\left(
(S^0)^{n}
\begin{vmatrix}
S^0_\l & S^1_\l\\
S^0_x & S^1_x
\end{vmatrix}
\right)_+\partial_y \mathbf{S} -
\nn\\
\left(
(S^0)^{n}
\begin{vmatrix}
S^0_x & S^1_x\\
S^0_y & S^1_y
\end{vmatrix}
\right)_+\partial_\l \mathbf{S}.
\label{Dun21}
\eea
It is easy to check that for $S^0=\lambda$ Dunajski equation hierarchy reduces
to heavenly equation hierarchy
\cite{Takasaki,Takasaki1}, while for $S^2=y$ it reduces to dispersionless
KP hierarchy.
\subsection{Waterbag-type reduction}
Discussing a special solution of Dunajski equation,
we have considered a reduction (\ref{firstred}) characterized by the existence
of two analytic solutions of linear equations (\ref{Lax2}).
Now we will introduce waterbag-type reduction
\beaa
&&
S^0=\lambda+\sum_{n=1}^N\ln\left(\frac{\l-u^0_n}{\l-v^0_n}\right),
\\&&
S^1=\sum_{n=0}^\infty t^1_n (S^0)^{n}+
\sum_{n=1}^N\ln\left(\frac{\l-u^1_n}{\l-v^1_n}\right)
\\&&
S^2=\sum_{n=0}^\infty t^2_n (S^0)^{n}+
\sum_{n=1}^N\ln\left(\frac{\l-u^2_n}{\l-v^2_n}\right),
\eeaa
where the functions $u$, $v$ depend only on times of the hierarchy.
This anzats is consistent with equations (\ref{Dun1},\ref{Dun2},\ref{volume})
and defines (1+1)-dimensional reduction of Dunajski equation hierarchy.
The first equation of this hierarchy, containing only the variables $x$, $y$,
is obtained by the substitution of the anzatz to relation (\ref{volume}).
The expression for the determinant will have $6N$ simple poles, and
relation (\ref{volume}) will give a closed system of $6N$ equations
for $6N$ functions $u$, $v$.

\section*{Acknowledgment}
The authors were partially supported by Russian Foundation for
Basic Research under grant no. 06-01-90840 and grant no. 06.01 CRF
CSTD AS Moldova. LVB and SVM were also supported in part by RFBR grant 04-01-00508.


\end{document}